# Design space exploration for image processing architectures on FPGA targets

Chandrajit Pal, Avik Kotal, Asit Samanta, Amlan Chakrabarti, Ranjan Ghosh

University College of Science, Technology and Agriculture,
University of Calcutta,
92, APC Road, Kolkata, India
http://www.caluniv.ac.in/

*Abstract.* Due to the emergence of embedded applications in image and video processing, communication and cryptography, improvement of pictorial information for better human perception like de-blurring, de-noising in several fields such as satellite imaging, medical imaging, mobile applications etc. are gaining importance for renewed research. Behind such developments, the primary responsibility lies with the advancement of semiconductor technology leading to FPGA based programmable logic devices, which combine the advantages of both custom hardware and dedicated DSP resources. In addition, FPGA provides powerful reconfiguration feature and hence is an ideal target for rapid prototyping. We have endeavored to exploit exceptional features of FPGA technology in respect to hardware parallelism leading to higher computational density and throughput, and have observed better performances than those one can get just merely porting the image processing software algorithms to hardware. In this paper, we intend to present an elaborate review, based on our expertise and experiences, on undertaking necessary trans-formation to an image processing software algorithm including the optimization techniques that makes its operation in hardware comparatively faster.

Keywords: *IP(intellectual property), FPGA(Field Programmable Gate Array), non-recurring engineering costs (NRE), FPGA-in-the-loop (FIL).*

## 1 Introduction

Human beings have historically relied on their vision for tasks ranging from basic instinctive survival skills to detailed and elaborate analysis of works of art. Our ability to guide our actions and engage our cognitive abilities based on visual input is a remarkable trait of the human species, and much of how exactly we do what we intend to do and seem to do it so well remains to be explored. The need to extract information from images and interpret their contents is one of the driving factors in the development of image processing and computer vision for the past decades, which demands for processing of the same to extract use-ful information from it. Digital image processing (DIP) is an ever growing area with a variety of applications including medicine, video surveillance and many more. To implement the upcoming sophisticated DIP algorithms and to process the large amount of data captured from sources such as satellites or medical instruments, intelligent high speed real-time systems have become imperative [1]. Image processing algorithms implemented in hardware (instead of software) have recently emerged as the most viable solution for improving the performance of image processing systems. Our goal is to familiarize applications programmers with the state of the art in compiling high-level programs to FPGAs, and to survey the relevant research work on FPGAs. The outstanding features, which FPGAs o er such as optimization, high computational density, low cost etc, make them an increasingly preferred choice of experts in image processing eld today. Technological advancement in the manufacture of semiconductor ICs of-fers opportunities to implement a wider range of imaging operations in real time. Implementations of existing ones need improvement. With the intrusion of reconfigurable hardware devices together with system level hardware description languages further accelerated the design and implementation of image process-ing algorithm in hardware. Due to the possibility of ne-grained parallelism of imaging operations, FPGA circuits are capable of competing with other calculation based implementation environments. This advancement have now made it possible to design complete embedded systems on a chip (SoC) by combining sensor, signal processing and memory onto a single substrate. With the ideal use of System-on-a-Programmable-Chip (SOPC) technology FPGAs prove to be a very efficient, cost-effective and attractive methodology for design verification [2].

In this paper we survey the various hardware implementation of image processing algorithms and show how the DSP design environment from Xilinx can be used to develop hardware-based computer vision algorithms from a system level approach, making it suitable for developing co-design environments with an emphasis on the salient features of FPGA. Section 2 highlights the setback of other hardware implementation alternatives and serves to set the basis for explaining the advantage of FPGAs while dealing with and evaluating several significant parameters. Section 3 summarizes the related research on FPGA implementation of image processing algorithms. Section 4 deals with the main contributions of the Xilinx DSP design environment, with the application examples and hard-ware architectures, 5 deals with the results and discussion and finally section 6 concludes the work with the discussion and projection towards future work.

## 2   Software paradigm to hardware(FPGA)

In general, sophisticated image processing algorithms are so computationally intensive that general-purpose CPUs cannot satisfy real-time constraints [3]. Software provides the flexibility and re-programmability features but leads to sequential execution of instructions and also increases the compiler overhead capable of identifying and execution of multi-thread components. However execution in customized hardware is inherently parallel as of its architecture and as a result the independent instructions of the algorithm can be executed in parallel subject to the availability of suitable hardware components, thereby increasing the speed of execution. Gains are made in two ways, while comparing hardware implementation with a software counterpart.

Firstly, a software implementation is constrained to execute only one instruction at a time. Although the life cycle of the instruction fetch/decode/execute cycle may be pipelined, and modern processors allow different threads to be executed on separate cores, software is inherently sequential by nature. A hardware implementation, on the other hand is fundamentally parallel, with each operation or instruction implemented on separate hardware module. In fact a hardware system must be explicitly programmed to perform operations sequentially if necessary. If an algorithm can be implemented in parallel to efficiently make use of the available hardware, considerable performance gains can be achieved.

Secondly, a serial implementation is memory bound, with data communicated from one operation to the next through memory. As a result a software processor needs to spend a significant proportion of its time reading its input data from memory, and writing the results of each operation ( including intermediate operations ) to memory.

Traditional digital signal processors are microprocessors designed to perform a special purpose, are well-suited to algorithmic-intensive tasks but are limited in performance by clock rate and the sequential nature of their internal design. This limits the maximum number of operations per unit time that they can carry out on the incoming data samples. Typically, three or four clock cycles are required per arithmetic logic unit (ALU), which lead to lower throughput. Multicore architectures may increase performance, but these are still limited. Designing with traditional signal processors therefore necessitates the reuse of architectural elements for algorithm implementation. In order to increase the performance of a system the number of processing elements needs to be increased, which has a negative effect of shifting the paradigm of concentration from signal processing to task overhead in controlling multiple processing elements.

A solution to this increasing complexity of DSP ( Digital Signal Processing ) implementations ( e.g digital lter design for multimedia applications ) came with the introduction of FPGA technology, developed as a means to combine and concentrate discrete memory and logic, thus enabling higher integration, higher performance and increased flexibility with their massively parallel struc-tures containing a uniform array of configurable logic blocks ( CLBs ), memory, DSP slices along with other elements [4],[5].

Nevertheless with the constant advancement of semiconductor technologies, FP-GAs are becoming sufficiently more powerful to support real-time image processing due to their high logic density, generic architecture and considerable on-chip memory. Moreover, the straightforward reconfiguration procedure allows designers to configure the hardware as many times as needed without extra cost i.e the ability to tailor the implementation to match system requirements. With these benefits there is a continued hardware design to meet the vertical requirements to meet the time critical and computationally complex applications that can be achieved through FPGA. Moreover its very high-speed I/O further reduces cost and minimizes bottlenecks by maximizing data flow right from capturing through the processing chain to the nal output. Sometimes constant upgradation in the device is required where ASICs (Application Specific Integrated Circuits) doesn't t well, as once it is programmed it cannot be changed [6].

Most machine vision algorithms are dominated by low and intermediate level image processing operations, many of which are inherently parallel. This makes them amenable to a parallel hardware implementation on an FPGA, which have the potential to significantly accelerate the image processing component of a machine vision system.

On an FPGA system, each operation is implemented in parallel, on separate hardware component allowing data to pass directly from one operation to an-other, significantly reducing or even eliminating the memory overhead. Fortunately, the low and intermediate level image processing operations typically used in a machine vision algorithm can be readily parallelized. FPGA implementation results in a smaller and more significantly lower power design that combines the flexibility and programmability of software with the speed and parallelism of hardware [7].

Hence, we choose an FPGA platform to rapidly prototype and evaluate our design methodology.

### 2.1  Evaluating FPGA with its advantages and disadvantages as a platform suitable for digital image processing applications.

Benefits of FPGA:
There are several advantages that makes FPGA a preferred choice as it o ers a convenient and flexible platform where real time machine vision systems can be implemented.

- In general, various image processing algorithms require multiple iterative processing of data sets as will be elaborated in the subsequent sections, requires sequential operations on a general purpose computer with multiple passes. It can be fused to one pass in an FPGA. It can be operated on multiple image windows in parallel as well as multiple operations within one window also in parallel.
- Optimization techniques such as loop unrolling, loop fusion etc help to effectively utilize the FPGA resources while maintaining the proper acceleration by reducing many redundant operations.
- Any digital logic circuitry can be configured differently as per the need of the hour and application at hand. So rapid prototyping of the devices are possible, which helps to test any architectural design we need to perform in a short time to market. Its software like flexibility to reprogram and easy upgradeability allows its solutions to evolve quickly.
- FPGA's inherent parallel configurable components, parallel programmable I/O, allow them to read, process and write from memory banks simultaneously. As result operations such as convolutions, correlations, digital FIR filtering can be done much faster using pipelining and parallelism.

- This reconfigurable and reusability feature of FPGA helps to develop im-age processing IP CORES, thus helps to generate most cost effective smart systems. These IP's can be quickly integrated without any moderation or repeating any verification reduces the time to market and reduces the non-recurring engineering (NRE) costs.
- There is a high logic as well as computational density within the FPGA together with a low development metric allows the lowest volume consumer electronics market to bear the development cost of FPGA. They are useful for low volume applications unlike ASIC's.
- Since we use hardware description language for designing the RTL model, the flexibility and configurability of FPGA comes out of it together with the speed and parallelism, which comes from the hardware implementation [8].

Shortcomings of FPGA The limitations of FPGA as faced in image process-ing operations are noted below:

- Hardware supports inherent parallel operations as per their architecture, and as a result offers much greater speed than software execution. But at the cost of an increased development time and proper skill needed by a design engineer.
- As it is used for product prototyping, its timing path cannot be fixed and optimized in advance as it needs to be changed with programming. As a result it operates at a very lower clock speed unlike ASIC.
- Since they are general purpose and programmable, they require large chip (silicon) area and consume more power.
- With FPGA Floating point operations are cost effective and complex mathematical operations such as division and direct multiplication are also computationally expensive. So it remains a good choice for the designers to reformulate their algorithms to avoid complexity [9].

Nevertheless the advantages outnumber the limitations and FPGA will continue to be a preferable choice for the designer community for the days to come.

## 2.2   Algorithm to hardware design  flow

The work flow graph shown in Fig. 1 shows the basic steps of implementing an image processing algorithm in hardware. Step 1 requires a detailed algorithmic understanding and its subsequent software implementation. Secondly the design should be optimized from both the algorithm (e.g. using algebraic transforms) and hardware (using efficient storage schemes and adjusting fixed point computation specifications) viewpoints. Finally, the overall evaluation in terms of speed, resource utilization, and image fidelity, decides whether additional adjustments in the design decisions are needed. Once done FPGA-in-the-Loop Verification is carried out, which enables us to run the test cases faster. It also opens the possibility to explore more test cases and perform extensive regression testing on our designs ensuring that the algorithm will behave as expected in the real world. A good software design does not necessarily correspond to a good hardware design and this clearly serves the purpose as to follow the steps mentioned in Figure 1a.

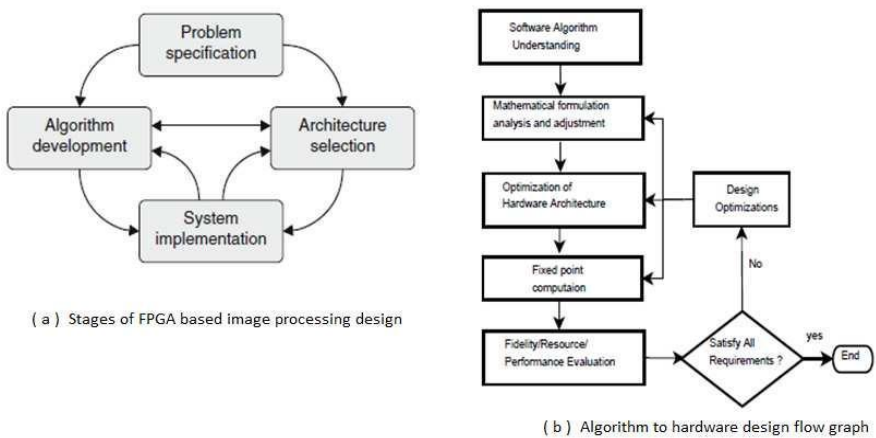

Fig. 1. Algorithm to hardware design flow graph.

## 3 Background and Related Work

Since 2000 we have seen a good amount of research on utilizing FPGA as a suit-able prototyping platform for realizing image and video processing algorithms. Digital image processing algorithms are normally categorized into 3 types: low, intermediate and high level. Low level operations are computationally intensive and operate on individual pixels and sometimes on its neighborhood involving geometric operation etc [7]. Intermediate-level operation includes conversion of the pixel data into different representation like histogram, segmentation, thresholding and the operations related to these. High level algorithms tries to extract meaningful information from the image like object identification, classification etc. As we move up from low to high level operations there is an obvious de-crease in the exploitable data parallelism due to a shift from pixel data to more descriptive and informative representations. Here we intend to focus on the low level operational (local filters) algorithms to deliberately show the capabilities of FPGA for computationally intensive tasks targeted for low and intermediate-level operations. As it is well known, a separate class of low level computationally intensive task includes image filtering operation based on convolution. Several related research works have been done so far.
Paper [10] have shown the various hardware convolution architectures related to look-up-table (LUT), distributed arithmetic and Multiplierless Convolution (MC) architecture and have stressed the usage of MC architecture since it is simple to implement and the multiplication operation can be replaced by an addition operation. However, such a realization is possible if only if a coefficient value is a power of 2 and is only favorable for small convolution kernels, thereby it loses its robustness. Paper [11] shows the various area efficient 2D shift-variant convolution architectures. They have proposed some novel FPGA-efficient architectures for generating a moving window over a row wise print path. Their moving window includes row major, column major and moving window with rotation stage architectures respectively. However their main architectural drawbacks is the memory overhead including an elevated memory bus bandwidth requirement as it needs to fetch multiple rows from external memory while processing a single row. Secondly more than one clock pulse is required for processing a single pixel. Paper [12] shows three different architectures for dealing with filter kernels whose coefficient value is varying. Their pipeline as well as convolve and gather architecture is worth noting. However they lag with some initial fixed redundant clock cycles used to buffer for the occurrence of the first convolution and an elevated pipelined architectural complexity, which comes from its construction of various segments meant for varying filter kernel coefficients.

Paper [13] discusses a multiple window partial buffering scheme for 2 dimensional convolutions. Their buffering strategy shows a good balance between on-chip resource utilization and external memory bus bandwidth suitable for low cost FPGA implementation. Paper [14] have shown an optimized implementation of discrete linear convolution. They have presented a direct method of reducing convolution processing time with computational hardware implementing discrete linear convolution of two finite length sequences. The implementation is advantageous with respect to operation, power and area optimization. Their claim that the architecture is capable of computing real time image processing algorithm for a particular application raises doubt since there is no validation results. Moreover for convolvers of large size it is recommended to use dedicated DSP blocks either as hard core or in software library while designing RTL for better performance issues.

Paper [15] shows the hardware architecture for 2D linear and morphological filtering applied to video processing applications. However video processing algorithm verification should not be done with USB, since it is much slower with respect to ethernet (point to point). Moreover they have used much slower clock frequency (10 MHz) to process, making it much unfamiliar.

## 4. Hardware convolution architectures

The convolution equation is given by

$$y[m,n] = x[m,n] * h[m,n] = \sum_{j=-a}^{a} \sum_{k=-b}^{b} x[j,k]h[m-j,n-k] \quad \text{--------- (1)}$$

where (m,n) are pixel positions, h[m,n] denotes the filter response function and x[m,n] is the image to be filtered. [a,b] denotes the window filter size [16].

The process scenario is clear from Fig.2.

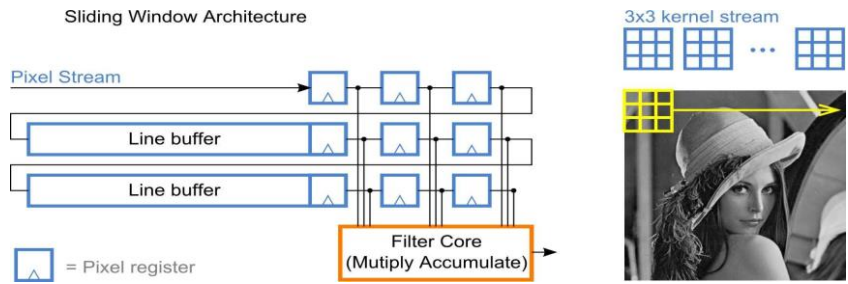

Fig. 2. Working procedure of a sliding window architecture.

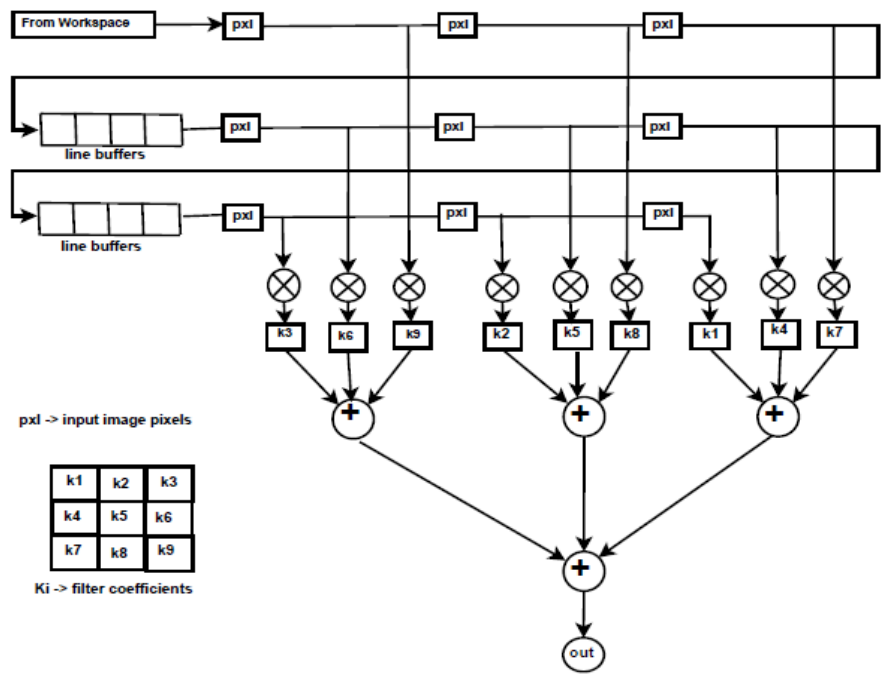

Fig. 3. Complete parallel hardware architecture of a 3x 3 filter kernel implementation for simplicity. Actually implemented 5x 5 kernel mask.

Here we have discussed five different convolution hardware architectures namely the fully parallel architecture, next an optimized version with MAC FIR lters, separable kernel architecture and another pipelined architecture capable of reducing some redundant operations. All of them have been designed to implement equation 1.

Fig.3 shows the buffer lines, which helps to store the image pixels prior to convolve, thereby saving additional time to fetch them from an external memory. Instead of sliding the kernel over the image this technique helps to feed the image through the window. This architecture is very common, which shows 2 buffer lines together with some memory registers, which assists in loading a 3*3 neighborhood. For the convolution operation it needs 9 multiplication and 8 addition operations and is a generic architecture with the highest complexity. This architecture computes a new output pixel at every clock cycle after an initial delay but consume more resources.

For Fig.4 The buffer line consists of a single port RAM, as shown in unit (2.a) of Fig. 4; the counter in it is incremented to write the current pixel data and to read it subsequently. The output of each of five buffers of unit-1 connects to respective inputs of unit-2, each of five parallel sub-circuits of unit-2 consists of five MAC FIR engines; one such unit is elaborately shown in unit-2.a of Fig. 4 depicting the ASR (Addressable Shift Register) implementing the input delay buffer. The address port runs n times faster than the data port, where n is the number of filter taps. The ROM and ASR address are produced by the counter. The sequence counts from 0 to n 1, then repeats. Pipeline registers r0 r2 increase performance. A capture register is required for streaming operation. A down sampler reduces the capture register sample period to the output sample period. The filter coefficients are stored in ROM. Five outputs of ve MAC engines are sequentially added to get the result, whose absolute value is computed and the data is narrowed to 8-bits. The blue colored block is elaborated in unit-2.b (Fig. 4) as the (multiply-accumulate)MAC engine. Enabling the 'Pipeline to Greatest Extent Possible' mask configuration parameter ensures the internal pipeline stages of the dedicated multipliers are used [17]. The yellow box is elaborated in unit 2.c (Fig. 4), which calculates the absolute value before multiplying with the scaling factor, which is the sum of the weight of the filter coefficients. This architecture has the advantage of using less resources but needs 5 clock cycles to process per pixel. The underlying 5-tap MAC FIR filters are clocked 5 times faster than the input rate. Therefore the throughput of the design is 100 Mhz/5= 20 million pixels per second. For a 64x64 image this is $20 \times 10^6/(64 \times 64)$= 4883 frames/sec. For our experiment the image size is 150x150, so 889 frames/sec. This architecture consumes very less hardware resources.

For linear operation, convolution has some interesting properties such as commutatively. Therefore for PxP kernels can be rede ned as the convolution of a Px1 kernel (Q1) with a 1x P kernel (Q2). As a result the equation can be formulated as

$$I \times Q1 \times Q2 = I \times Q2 \times Q1 \quad (2)$$

Fig.5 and 6 implements the right hand and left hand side of the equation 2 respectively. The design with separable convolution kernel architecture is shown in Fig. 5 and Fig.6. In Fig.5 the column convolution has been carried out in the rst section of the hardware before the row buffering scheme. The row bu ering is shown in the detailed architecture in unit 1.a of Fig.4 as explained previously and the row convolution in unit 4.a of Fig. 4 respectively. The partially processed pixels after the column convolution is passed through the row convolution section to get the filtered pixel and is capable of processing $(100 \times 10^6)/256 \times 256 = 1526$ frames/sec. 100 stands for the frequency of the FPGA board in MHz and image size is 256 x 256 and $100 \times 10^6/(150 \times 150) = 4444$ frames/sec for a 150x150 size image.

This architecture is capable of processing 1 pixel/clock cycle and its complexity is reduced from $O(N^2)$ for normal convolution as discussed to $O(2N)$.

Fig.7 takes the advantage of only five multiplications and two 4-operand additions. In other words this architecture reduces these redundant operations. But in contrast, this architecture has three mult-add pipelines, which allows to operate with three mask columns. It is to be noted that this architecture selects (to the output adder) 5-predefined input operands (see connections of inputs of this adder in Fig.7). This architecture also processes 1 pixel/clock cycle.

It is to be noted that the architecture shown in Fig.4 needs 5 clock cycles to process 1 pixel as shown in the timing diagram in Fig.8. The rest of all architectures in Figures 3, 5, 6 and 7 processes 1 pixel/clock cycle as shown in the timing diagram in Fig.12, 9, 10, 11.

For the above architectures discussed in section 4, the hardware resource utilization has been shown in Table 1.

## 5   Results and Timing Diagram

The corresponding hardware architectures have been applied for verifying an edge preserving bilateral filter, which involves execution of multiple convolution operations in parallel pipelining fashion. The results of the denoised image are as shown in Fig.13 and 14. Filter output for image size of 150x150 for the additive Gaussian noise. Filter settings $\sigma_s=20$, $\sigma_r=50$ and $\sigma=12$ for the additive Gaussian noise, where $\sigma_s$ and $\sigma_r$ are the domain and range kernel standard deviations and only $\sigma$ is the needed for the white Gaussian noise.

There remain some considerations while planning to implement complex image processing algorithms in real time. One such issue is to process a particular frame of a video sequence within 33 ms in order to process with a speed of 30 (frames per second) fps. In order to make correct design decisions a well known standard formula given by:

$$t_{frame} = C/f = (M.N/t_p + \xi)/n_{core}.f \leq 33ms \quad (3)$$

where $t_{frame}$ is the processing time for one frame, C is the total number of clock cycles required to process one frame of M pixels, f is the maximum clock frequency at which the design can run, $n_{core}$ is the number of processing units, $t_p$ is the pixel-level throughput with one processing unit ($0 < t_p < 1$), N is the number of iterations in an iterative algorithm and is the overhead ( latency ) in clock cycles for one frame [3].

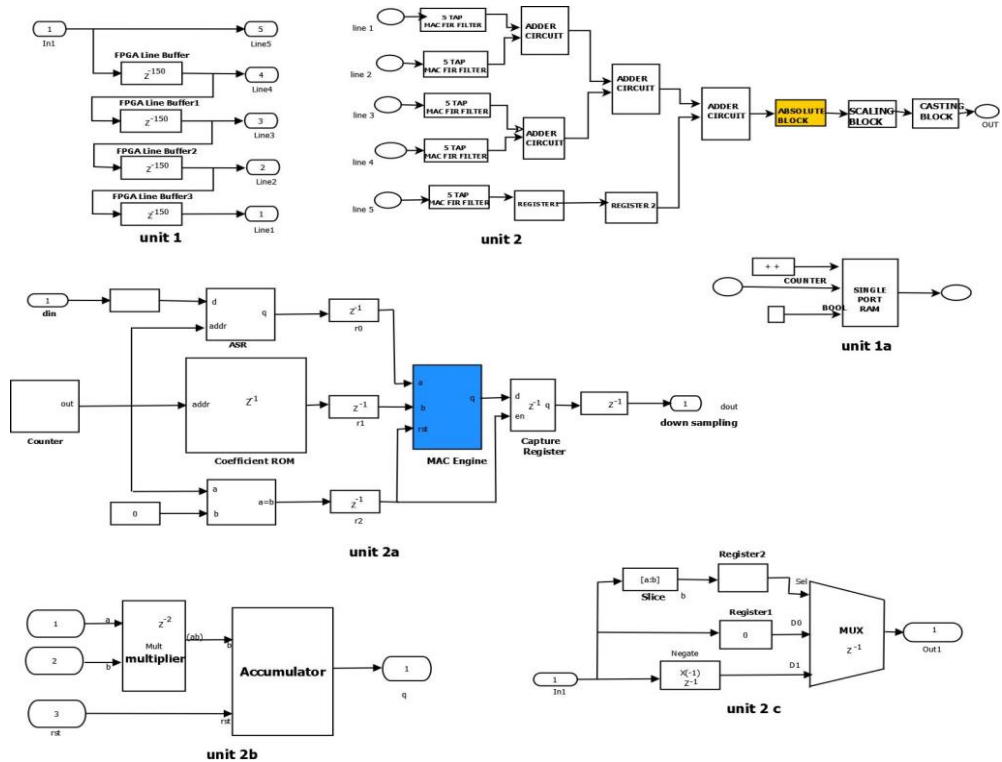

Fig. 4. Hardware blocks showing the ltering hardware architecture of a 5x5 filter kernel implementation [18].

We have tested for our convolution architectures discussed above for a single image filtering application and have measured the time via the well known eqn 3 [3].

For 150 x 150 resolution image, M= 22500, N = 1, $t_p$ = 1 i.e per pixel processed per clock pulse, and = 350 i.e the latency in clock cycle, f = 100 MHz, $n_{core}$ = 1. Therefore the $t_{frame}$ = 0.00022 seconds = 0.2 ms 33ms ( i.e much less than the minimum timing threshold required to process per frame in real time video rate ). We have measured the same execution in software and it came to be 0.008 second. Therefore the acceleration in hardware is 0.008/0.00022 = 40x . From Table 1 it is clear that architecture in Fig.5, 6 and 7 are most suitable w.r.t resource usage. We have also measured the power consumption of the individual hardware architectures as shown in Table 2. From the data it is clear that the normal convolution hardware in Fig.4 and the separable hardware architectures in Fig.5, 6 consumes the least power among the rest.

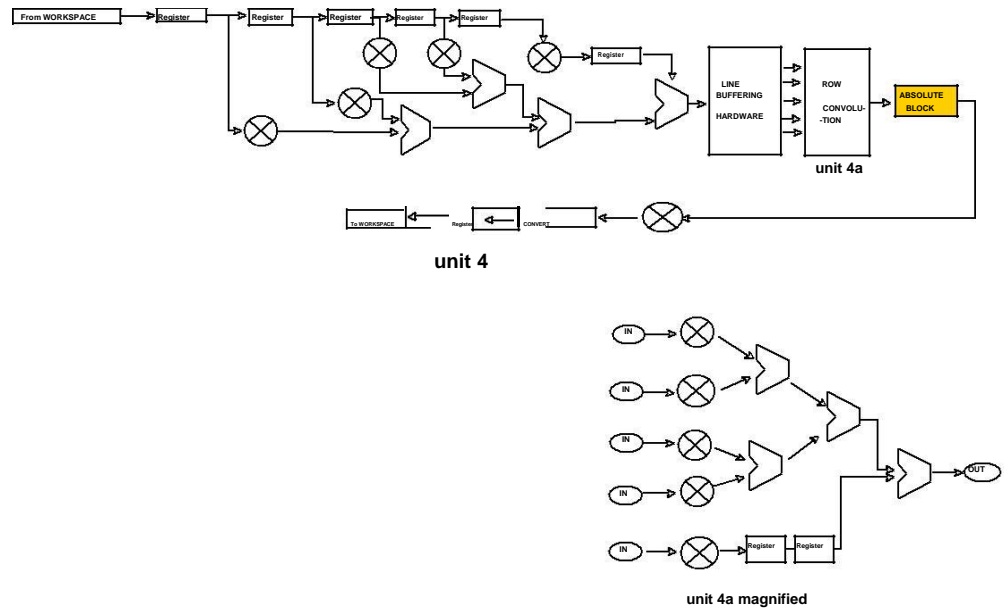

Fig. 5. Hardware blocks showing the filtering hardware architecture for separable kernel. Right hand side of Eqn. 2.

## 6   Discussions and Future Directions

In this paper we have discussed in brief our motivation towards the computer vision algorithm implementation realized in hardware and presented various e efficient convolution architectures with almost similar results, with minute changes in the PSNR of the filtered output images resulted after applying Gaussian filtering on a noisy image shown in Fig.13. We have also tested our architectures, which when applied to a particular edge preserving algorithm produced good results (with enhanced PSNR as shown in Fig.13). It has been shown that Xilinx System Generator (XSG) environment can be used to develop hardware-based computer vision algorithms from a system level approach, making it suitable for developing co-design environments. We have also used FPGA-in-the-loop (FIL) verification [19], to verify our design. This approach also ensures that the algorithm will behave as expected in the real world. In future we need to explore more high level technique and approaches to circuit optimization with energy efficiency.

Table 1. DEVICE UTILIZATION OF THE VARIOUS OPTIMIZED HARD-WARE ARCHITECTURES FOR IMAGE SIZE 150x150 FOR VIRTEX 5 LX110T OpenSPARC EVALUATION PLATFORM

| Percentage utilization | Image Size (150x150) | | | |
|---|---|---|---|---|
| | Normal Convolution hardware(Fig.4) | fully parallel architecture(Fig.3) | SSDC hardware (Fig.5 and 6) | architecture in Fig.7 |
| occupied slices out of 17,280 | 525 (4%) | 1586 (9%) | 623 (4%) | 740 (4%) |
| Slice LUTs out of 69,120 | 1062 (2%) | 2922 (4%) | 1593 (3%) | 1595 (2%) |
| Block-RAM/FIFO out of 148 | 7 (5%) | 6 (4%) | 6 (4%) | 6 (4%) |
| Flip Flops out of 69,120 | 4041 (6%) | 4042 (6%) | 810 (2%) | 1890 (3%) |
| IOBs out of 640 | 1 (1%) | 1 (1%) | 1 (1%) | 1 (1%) |
| Mults/DSP48s out of 64 | 5 (8%) | 0 (0%) | 0 (0%) | 0 (0%) |
| BUFGs/BUFCTRLs out of 32 | 2 (6%) | 2 (6%) | 2 (6%) | 2 (6%) |
| *SSDC = Separable Single Dimensional Convolution | | | | |

Table 2. POWER CONSUMPTION OF THE VARIOUS OPTIMIZED HARD-WARE ARCHITECTURES FOR IMAGE SIZE 150x150 FOR VIRTEX 5 LX110T OpenSPARC EVALUATION PLATFORM

| Power Consumption | Image Size (150x150) | | |
|---|---|---|---|
| | Static Power (in Watt) | Dynamic Power (in Watt) | Total Power (in Watt) |
| Normal Convolution Hardware in Fig.4 | 0.703 | 0.041 | 0.744 |
| Separable Hardware architecture in Fig.5,6 | 0.702 | 0.025 | 0.728 |
| Architecture in Fig.7 | 1.188 | 0.072 | 1.26 |
| Fully Parallel arch. Hardware in Fig.3 | 1.188 | 0.068 | 1.26 |

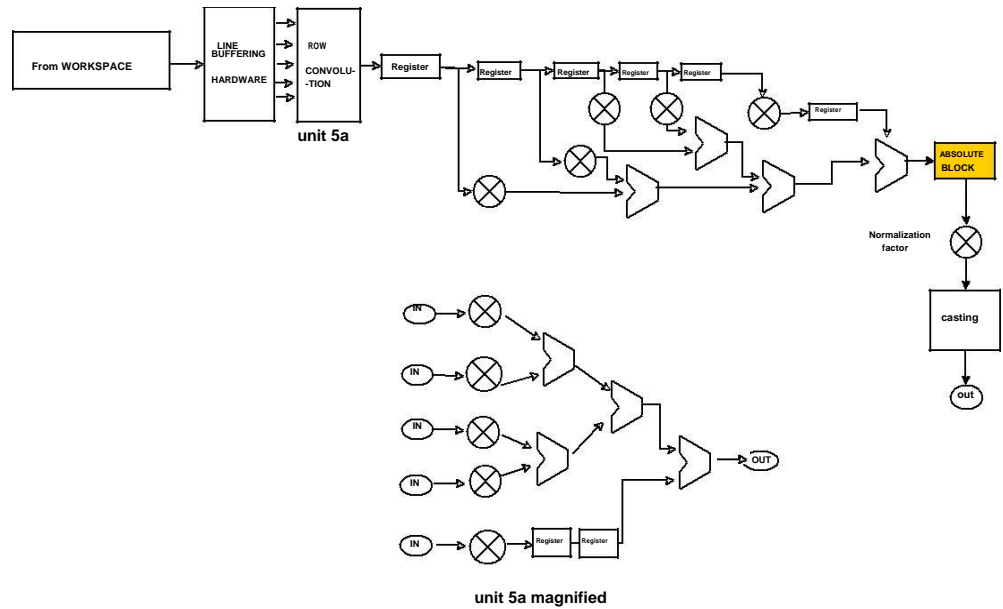

Fig. 6. Hardware blocks showing the filtering hardware architecture for separable kernel. Left hand side of Eqn. 2.

## Acknowledgment

This work has been supported by the Department of Science and Technology, Govt of India under grant No DST/INSPIRE FELLOWSHIP/2012/320 as well as grant from TEQIP phase 2 (COE), University of Calcutta for the experimental equipments. The authors wish to thank Dr. Kunal Narayan Chaudhury for his help regarding some theoretical understandings.

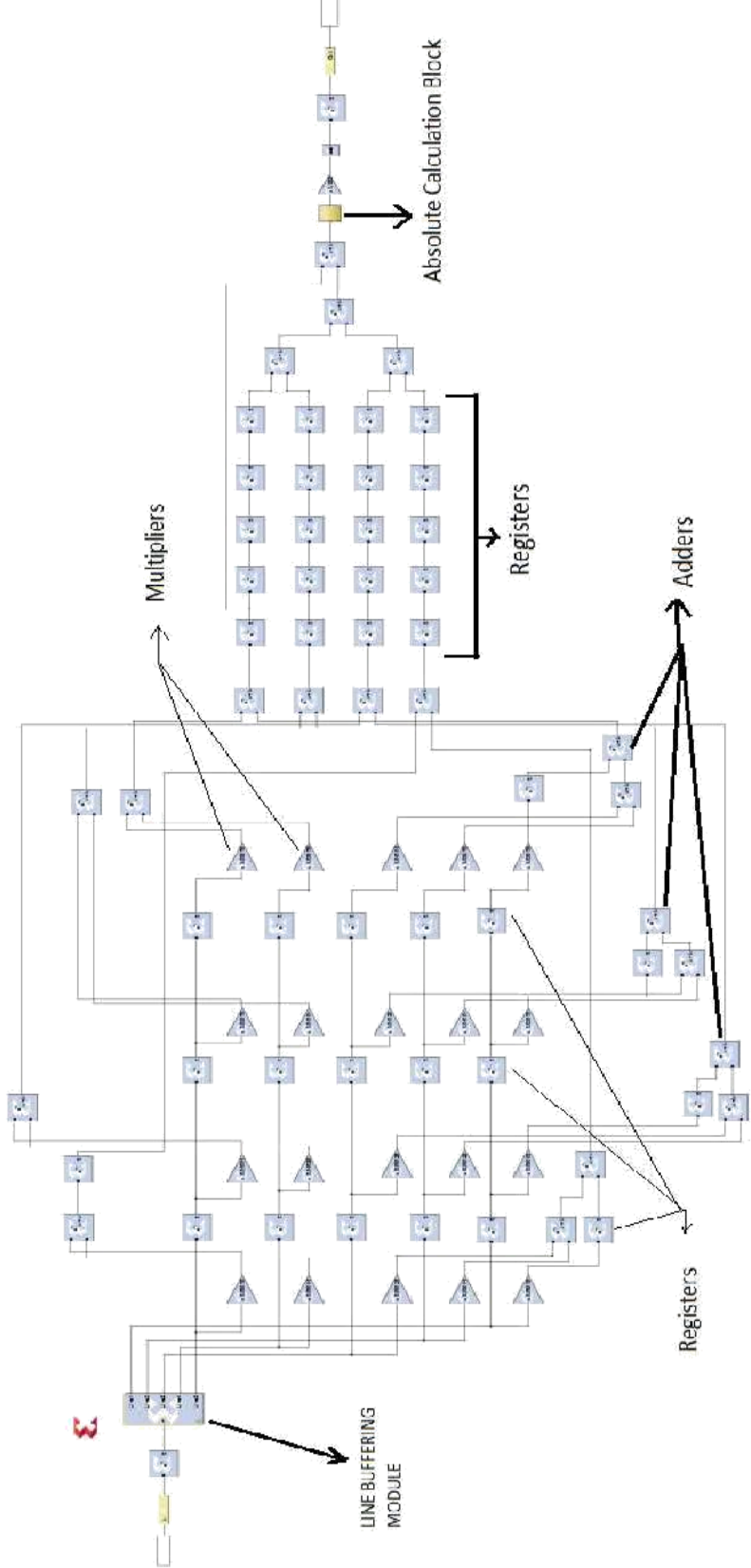

Fig. 7. An optimized convolution architecture developed to work with kernels like Gaussian, high pass filters, point and line detection etc.

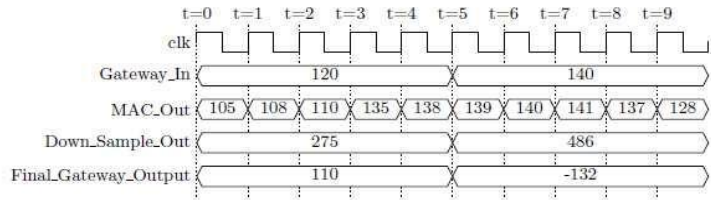

Fig. 8. Simulation results showing the time interval taken to process the image pixels for a normal convolution hardware architecture in Fig.4 where 5 clock pulses are needed to process per pixel. Each clock pulse duration is 10 ns.

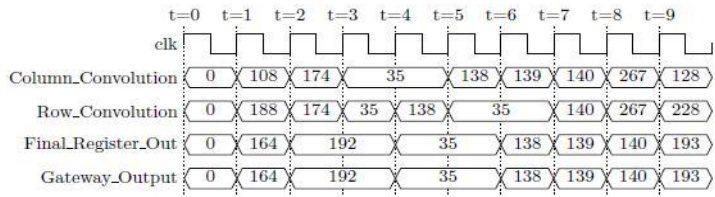

Fig. 9. Simulation results showing the time interval taken to process the image pixels. Each clock pulse duration is 10 ns. Each pixel requires one clock pulse to process. This timing diagram is followed by all the architectures except for Fig.4. It is implementing right hand side of equation 2.
.

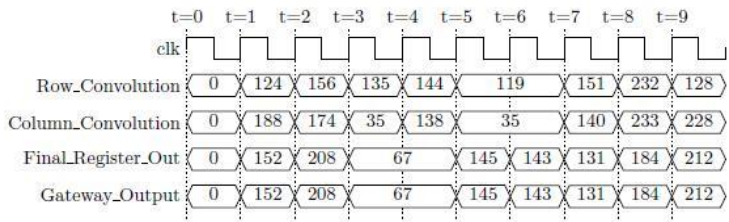

Fig. 10. Simulation results showing the time interval taken to process the image pixels. Each clock pulse duration is 10 ns. Each pixel requires one clock pulse to process. This timing diagram is followed by all the architectures except for Fig.6. It is implementing left hand side of equation 2.
.

 

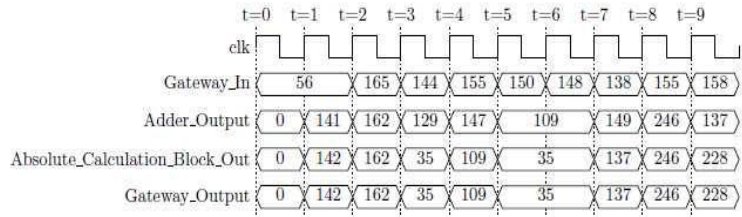

Fig. 11. Simulation results showing the time interval taken to process the image pixels. Each clock pulse duration is 10 ns. Each pixel requires one clock pulse to process. This timing diagram is followed by all the architectures except for Fig.7.
.

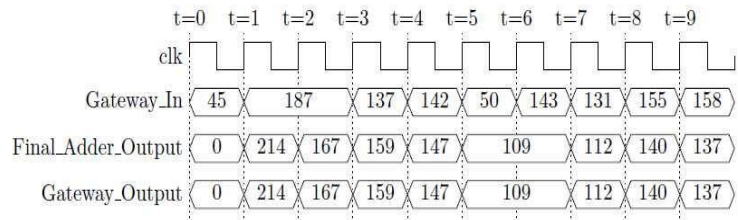

Fig. 12. Simulation results showing the time interval taken to process the image pixels. Each clock pulse duration is 10 ns. Each pixel requires one clock pulse to process. This timing diagram is followed by all the architectures except for Fig.3 and it is a complete parallel architecture.
.

   

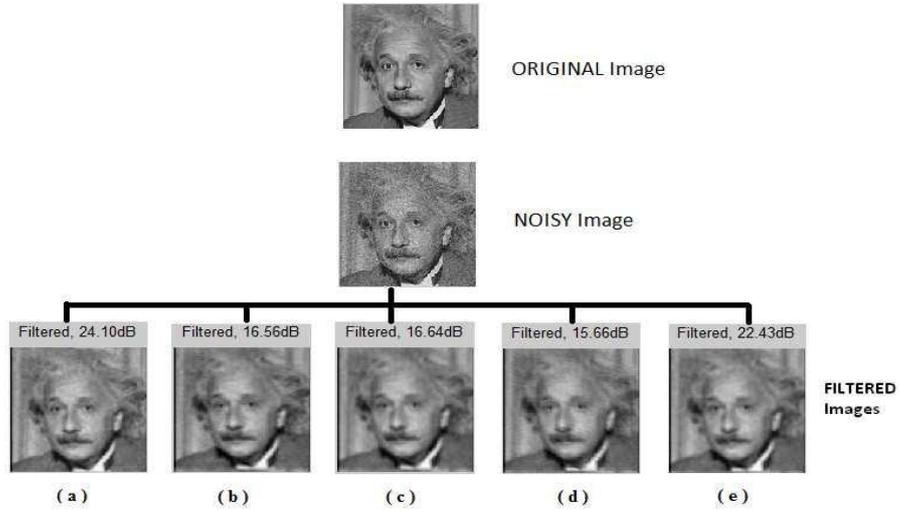

Fig. 13. Gaussian filtered output for image size of 150 150 applied over noisy image with (variance) $\sigma^2$ = 0:005. Filter settings $\sigma_s$=20 (domain kernel std dev). The filtered images (a),(b),(c),(d) and (e) correspond to the architectures shown in Figures 4, 5, 6,7 and 3.

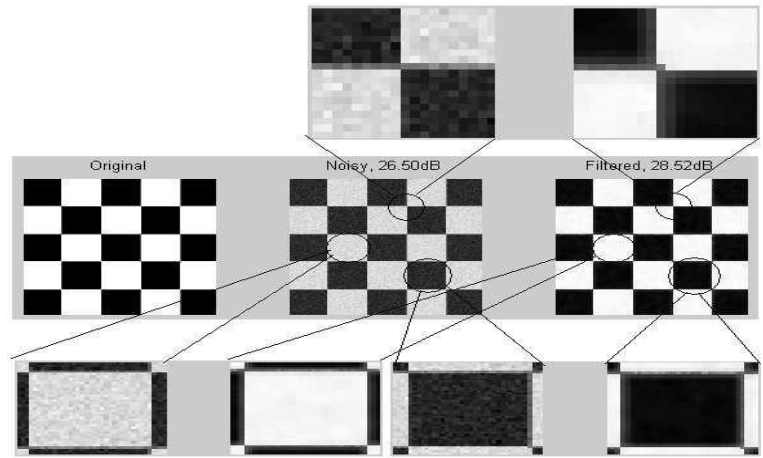

Fig. 14. Filter output for checkerboard image of size 150x150 for the additive Gaussian noise. Filter settings $\sigma_s$=20, $\sigma_r$=50 and $\sigma$ =12 for the additive Gaussian noise [18].

11. Cardells-Tormo, F. Molinet, P, for Area-e cient 2-D shift-variant convolvers FPGA-based digital image processing, IEEE Workshop on Signal Processing Systems Design and Implementation, 2005, pp:209-213, doi: 10.1109/SIPS.2005.1579866.